# Effects of Diffusion on Photocurrent Generation in Single-Walled Carbon Nanotube Films

Christopher A. Merchant and Nina Marković

*Department of Physics and Astronomy, Johns Hopkins University*

*Baltimore, MD 21218*

We have studied photocurrent generation in large carbon nanotube (CNT) films using electrodes with different spacings. We observe that the photocurrent depends strongly on the position of illumination, with maximum observed response occurring upon illumination at the electrode edges. The rate of change of the response decays exponentially, with the fastest response occurring for samples with the smallest electrode spacing. We show that the time response is due to charge carrier diffusion in low-mobility CNT films.

Carbon nanotubes have been considered for incorporation into photovoltaic devices because of their unique electronic and mechanical properties[1,2]. CNT/polymer composites and transparent CNT films have been studied as electrodes[3-5] and photocurrent collectors[6,7] for such devices. Semiconducting CNTs themselves hold promise as photocurrent generators[8,9] because they possess a bandgap suitable to visible-light wavelengths[2]. It is believed that the Schottky barrier at the CNT-metal interface is responsible for the electron-hole separation necessary for photocurrent production from CNTs[10,11]. While the photocurrent time response of single CNTs has been observed to be expectedly fast[10,12], the origin of the very slow time response of CNT films[13-16] has been debated. In this work, we have studied the photocurrent generation in large CNT films with different electrode spacings. We show that the slow time response of CNT films is a result of charge carrier diffusion in a low mobility medium[17].

CNT photovoltaic devices were created using a spray-coating technique[18]. A schematic of a typical device is shown in Figure 1(a). Two Au electrodes were deposited on glass slides via thermal evaporation. The spacing between the electrodes ranged from 0.5 to 27 mm. Commercially-obtained SWNTs (buckyUSA) were then dispersed in 1,2-dichloroethane and sprayed on top of the electrodes using an airbrush system. A photograph of a sample is given in Fig. 1(b).

Sample resistance as a function of the amount of CNTs (in mg) is shown in Fig. 1(c) for multiple devices. As the CNT film thickness $s$ increases linearly with the mass $m$ of applied CNTs, the resistance of the films should be inversely proportional[19] to $m$, as $R \sim s^{-1} \sim m^{-1}$. We indeed find that the resistance varies as $R = d\alpha m^{-1}$, where $\alpha$ is a constant and $d$ is the spacing between electrodes. Our measurements yield $\alpha = 140$ Ω·mg/mm.

The CNT films were illuminated with a broadband fiber-optic light (Dolan-Jenner 190) to prevent sample heating. The light source was positioned relative to the sample so that incident light spot size is 0.5 cm$^2$. Photocurrent was measured during illumination without the application of a bias voltage. The dark current[20], the current observed without illumination, has been subtracted from all of the photocurrent measurements. Figure 2(a) shows the steady-state photocurrent, $I_0$, as a function of the position of illumination along one representative sample. Similar behavior has been observed in all of our samples. Samples were illuminated along the axis indicated in the inset of Fig. 2(a). Measurements begin in the center of the positive electrode and continue at 1mm intervals, until the center of the negative electrode is reached. As shown in Fig. 2 (a), the amplitude of the photocurrent response is greatest when illuminating directly over the edges of the electrodes. We also find that the overall current level increases as we add CNT layers. For the particular sample shown, the maximum current level at the negative electrode edge increased from 3.7 µA to 8.2 µA through the application of 3 extra CNT layers. In this same range, the sample resistance was observed to drop from 200 Ω to 33 Ω.

From Fig. 2(a) we see that the absolute value of the photocurrent peaks at edges of the positive and negative electrodes, but decreases again as the light spot moves into the center of the electrode. A 'position effect' has been observed in CNT photovoltaic devices[14-16], however, the decrease in the photocurrent for illumination at the center of the electrodes has not been observed previously. This is because in previous studies the light spot size was of the same order of magnitude as the electrodes.

Figure 2(b) shows the time response of the sample in Fig. 2(a), upon illumination, for different positions. We see that the time response of the photocurrent, I(t), is described well by an exponential $I(t) = I_0 \cdot (1-exp(-t/\tau))$ where $I_0$ is the steady-state photocurrent plotted in Fig. 2(a) and $\tau$ is a time constant.

The exponential time response of the photocurrent to a step-function illumination is shown for two samples with different electrode spacing in Fig. 2(c). The time constant is clearly larger for the sample with larger electrode spacing. For these measurements the samples were illuminated at the edge of the positive electrode using a 1mm×5mm light spot. The value of $\tau$ ranged between 0.3 seconds and 5.5 seconds for all of our measurements on samples with electrode spacing from 0.5 to 27 mm. While these responses are much slower than expected for raw photoproduction of electrons from CNTs[21], they are of the same order as values observed by other groups for CNT bundles and mats[13-15].

There has been considerable debate regarding this very slow time response of CNT-metal heterojunctions. Suggestions to explain the slow response have included oxygen desorption[22] and thermal effects[16]. In contrast, we show that the slow time response can be explained using a model of charge-carrier diffusion. We start by assuming that the illumination of the samples produces electron-hole pairs in the CNT film[8]. As some energetic electrons are able to cross the Schottky barrier at the CNT-metal interface, they leave behind an excess of holes in the vicinity of the electrode. This excess of holes retards photocurrent generation as it leads to increased recombination of photoproduced electrons. The maximum photocurrent will be realized only when the extra holes have diffused away from the area of illumination towards the other electrode. The time response for the photocurrent to be maximized would, therefore, depend on the distance between electrodes and on the hole mobility in the CNT films.

The excess hole density distribution between the electrodes should depend on how the physical width of the illumination source (~1mm) compares with the electrode spacing. We expect the excess hole density distribution to go from a near-uniform distribution (when the electrode spacing ≈ light spot size) to a quickly decaying distribution (when the electrode spacing >> light spot size). This can be described by a parabolic density distribution of the form $N(x) = N_0/d^r(d-x)^r$, where $d$ is the electrode spacing, $x$ is the position between the electrodes, and $r$ is an exponent. This type of parabolic distribution has been used before to describe the donor impurities in the base region of a transistor[23,24]. In Fig. 3(a) we show the parabolic density distributions for three different values of $r$.

We expect that as the electrode spacing increases, the parameter $r$ should increase monotonically. Figure 3(b) shows a trial function of the parameter $r$ against the electrode spacing $d$. We have used a power-law form for $r$ such that $r \sim d^a$, where $d$ is the electrode spacing, and $a$ is a fitting parameter. The trial form plotted in Fig. 3(b) uses a parameter $a = 1.8$, and varies from $r = 0$ ($d<1$) to $r \sim 100$ ($d\sim30$).

Assuming that the sample is illuminated at the edge of one electrode, the excess holes will diffuse towards the other electrode. The transit time can be calculated using a diffusion model for charge carriers, taking into account the recombination and the variable built-in electric field[24]. Using this model together with the parabolic hole density distribution, $N(x,d)$, the time constant can be calculated as:

$$\tau = \frac{d^2}{2D(r+1)}\left[1 - \frac{I_{\frac{1}{2}(r+3)}(d/L_p)}{I_{\frac{1}{2}(r-1)}(d/L_p)}\right] \qquad (1)$$

Here $I_n$ is a modified Bessel function of the first kind of order $n$, $d$ is the electrode spacing, $L_p$ is the diffusion length, and $D$ is the diffusion coefficient. By adjusting the form of the parameter $r$ in Fig. 3(b), we can obtain the time constant for samples with varying electrode spacing.

Figure 3(c) shows the measured time constant as a function of the electrode spacing along with the calculated values from Eq. 1. To calculate $\tau$, we used the fitting parameters $L_p$ = 1mm and $D$ = 0.01 cm$^2$/V·s. We are able to check the validity of our model by using this diffusion constant to calculate other physical properties of the film. The diffusion constant, $D$, is proportional to the mobility through the Einstein relation, $D = \mu k_B T/q$, where $k_B$ is the Boltzmann constant, $T$ is the temperature, $q$ is the electric charge, and $\mu$ is the mobility. As our measurements are conducted at room temperature, we calculate the mobility to be $\mu$ = 0.4 cm$^2$/V·s. This value is much lower than the expected value[25] for single CNTs of 100,000 cm$^2$/V·s, but is in the expected range of values for a CNT film[26]. Indeed, our value of the mobility is in close agreement with previous measurements for spray-formed CNT films[27]. We further check our model by estimating the charge carrier concentration, $n_e$. From Fig 1(c), we measure the resistivity of $\rho \approx$ 2 Ω·cm for a 13 mm long sample. Our calculated mobility, therefore, gives an estimate of $n_e \approx 6 \times 10^{-20}$ cm$^{-3}$. This is a fairly large carrier concentration, but within the expected values for SWNTs[28]. We conclude that the CNT films have high carrier concentrations as expected from the individual CNTs, with mobilities lower than that for single CNTs due to the nature of the random network.

In summary, we have demonstrated that the charge carrier diffusion in addition to the low mobility in CNT films appears to be responsible for the very slow time response associated with the photocurrent generation. Our conclusions could have a large impact on the design of CNT-based solar cells and photosensors, in terms of device geometry and the choice of materials.

This work is supported by in part by the National Science Foundation under Grants No. ECCS-0403964 and No. DMR-0520491 (MRSEC), Alfred P. Sloan Foundation under grant No BR-4380, and ACS PRF No 42952-G10.

**Figure 1:** a) Device schematic. Metal electrodes are evaporated onto a glass slide and the CNT films are added by spray-coating the entire sample. b) Photograph of a typical sample. c) Resistances of samples with electrode spacing of 1.5mm (open red squares), 5mm (open blue squares), 13mm (green x's), and 19mm (open purple triangles) are measured as a function of the amount of the applied CNT material. Fits are given by dashed black lines.

**Figure 2:** a) Maximum photocurrent as a function of position. Maximum photocurrent, $I_0$, is plotted as a function of position for CNT films of different thicknesses. Total amounts of CNTs in each layer were 2.5 mg (red cross), 7.5 mg (blue open square), 12.5 mg (green open circle), and 15 mg (black x). Position of the electrode edges are indicated by vertical dashed lines. Inset: The photocurrent measurement schematic shows where the samples were illuminated. Photocurrent measurements begin when the sample is illuminated in the center of the negative electrode (-10 mm position) and continue in 1mm intervals until the center of the positive electrode is reached (10 mm position). Measurements occur along the center of the glass slide (in the horizontal direction in the schematic). b) Time response of the photocurrent. Time response of the photocurrent is measured at different positions as in (a). Fits to the data (black dashed lines) are exponential and the positions of the measurements (x values) are indicated. c) Photocurrent as a function of time for samples with different electrode spacing. Samples with spacing of 1.5 mm (green cross) and 27 mm (red circle) are shown along with exponential fits (dashed lines) of the same form as in Fig. 2(b).

**Figure 3:** a) Carrier density between the electrodes. Density, N(x), is of a parabolic form with exponent, r. Densities are plotted for several exponents, r = 1, 3 and 10. b) Trial form of the exponent r as a function of electrode spacing. c) Minimum time constant, $\tau_{min}$, as a function of electrode spacing. Black squares are data and dashed red line is a fit using the exponent r in Fig. (b) and Eq. 1.

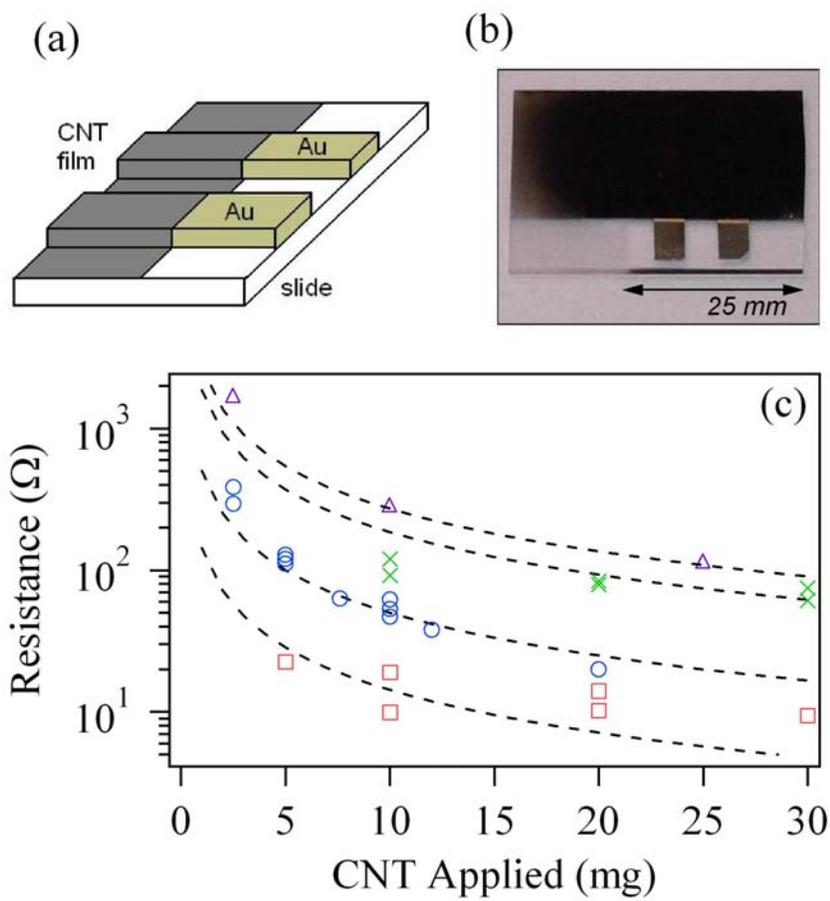

**Figure 1**

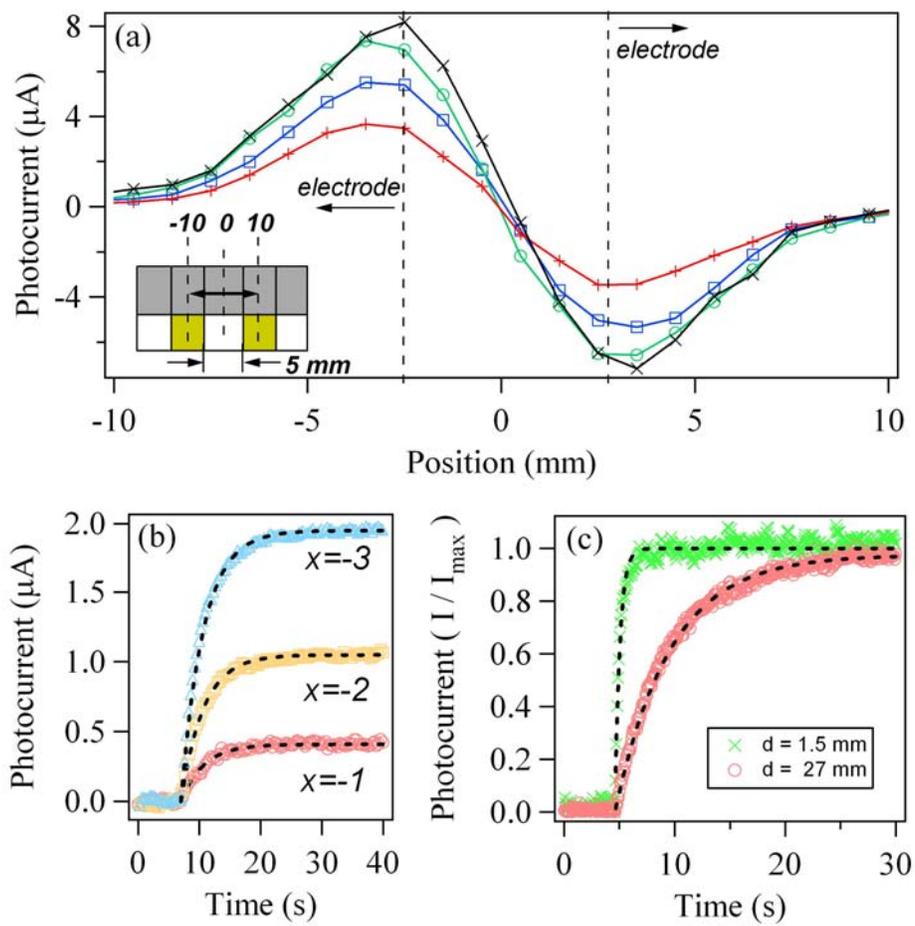

**Figure 2**

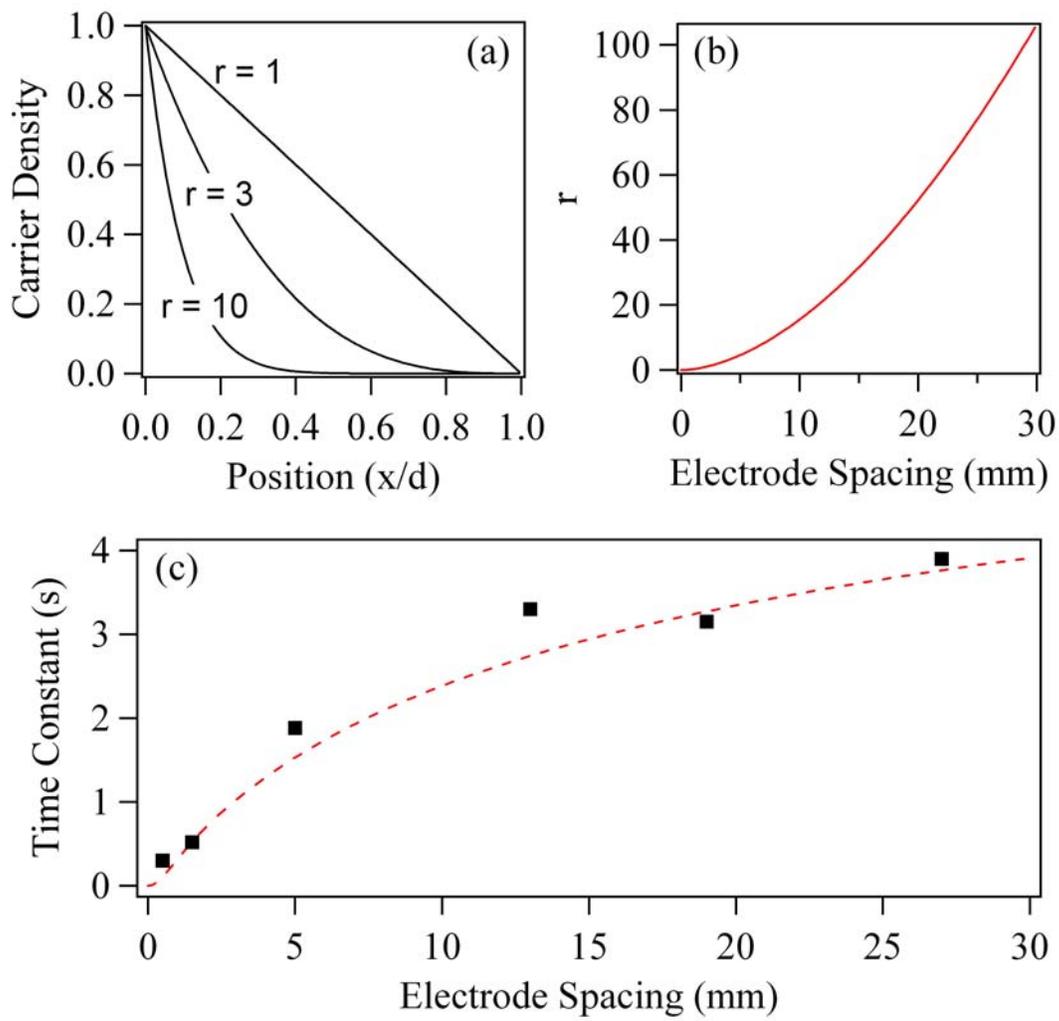

**Figure 3**